\documentclass[11pt,a4paper]{article}

\usepackage{amsmath}
\usepackage{amsfonts}
\usepackage{amssymb}
\usepackage{graphicx,psfrag,color}
\usepackage{subfigure}
\usepackage{hyperref}

\usepackage{enumerate}

\usepackage[left=2cm,top=2.5cm,right=2cm,bottom=2cm]{geometry}

\begin{document}
\noindent \textbf{Mod. Phys. Lett. A, Vol. 29, No. 25 (2014) 1450119}
\begin{center}
\Large{\textbf{Probing the matter and dark energy sources in a viable Big Rip model of the Universe}} \\[0.2cm]
 
\large{\textbf{ Suresh Kumar\footnote{E-mail: sukuyd@gmail.com}}}
\\[0.1cm]

\small{
\textit{ Department of Mathematics, BITS Pilani, Pilani Campus, Rajasthan-333031, India.}}

\end{center}

\vspace{.2cm}

\noindent \small{\textbf{\large{Abstract:}} Chevallier-Polarski-Linder (CPL) parametrization for the equation of state of dark energy in terms of cosmic redshift or scale factor have been frequently studied in the literature. In this study, we consider cosmic time based CPL parametrization for the equation of state parameter of the effective cosmic fluid that fills the fabric of spatially flat and homogeneous Robertson-Walker spacetime in general relativity. The model exhibits two worthy features: (i) It fits the observational data from the latest $H(z)$ and Union 2.1 SN Ia compilations matching the success of $\Lambda$CDM model. (ii) It describes the evolution of the Universe from the matter-dominated phase to the recent accelerating phase similar to the $\Lambda$CDM model but leads to Big Rip end of the Universe contrary to the everlasting de Sitter expansion in the $\Lambda$CDM model. We investigate  the matter and dark energy sources in the model, in particular, behavior of the dynamical dark energy responsible for the Big Rip end of Universe.\\

\noindent \small{\textbf{{PACS:}} 95.35.+d, 95.36.+x, 98.80.Cq
\section{Introduction}
\label{Intro}

The standard $\Lambda$CDM (cosmological constant + cold dark matter) model of Big Bang cosmology fits the observational data very well, and has been remarkably successful in describing the real Universe \cite{PlanckXXVI}. This model successfully describes the evolution of the Universe from the matter-dominated phase to the recent dark energy dominated and accelerated expansion phase in general relativity. It deserves mention that the cosmological constant $\Lambda$ is the simplest candidate of dark energy, which has constant energy density throughout the cosmic evolution \cite{8,paddy05,paddy05b,varun06}. And it suffers from theoretical problems such as fine tuning and cosmic coincidence \cite{10}. Also, it faces persistent challenges from observations on small scales known as ``small scale controversies", for instance,  the observations related to innermost regions of dark matter halos and the Milky Way dwarf galaxy satellites show the inconsistency with CDM paradigm \cite{vega11,weinberg13}. On the other hand, the nature of CDM itself is elusive, and there have been different approaches to understand its nature \cite{Rind14,Heik14,Kile13}. Thus, the $\Lambda$CDM model is plagued with number of problems despite its great success in describing the Universe. Further, the late $\Lambda$CDM Universe expands forever with the de Sitter phase.  However, we can not be sure about such a future of the real Universe because the observational data do not exclude the possibility of the domination of exotic dark energy stuff, which may lead to Big Rip end of the Universe \cite{PlanckXXVI,caldwell03,mel03,car03,Bennett13}.  This motivates us to search/construct the viable alternative models of the Universe, in particular the ones which offer a future of the Universe different from the de Sitter one, and fit the observational data matching the success of the $\Lambda$CDM model \cite{odi10,odi12a,odi12b,odi12c,odintsov13}.

The Chevallier-Polarski-Linder (CPL) parametrization of equation of state (EoS) parameter of dark energy  was first introduced in \cite{Chevallier01}, and reads as $w_{\rm de}=w_0+w_1(1-a/a_0)$, where $w_0$ and $w_1$ are constants, and $a$ is cosmic scale factor with $a_0$ being its present value. One may note that the CPL parameterization carries first two terms of Taylor expansion of $w_{\rm de}$ in terms of $a$ about $a_0$, and hence it  is naturally motivated, and approximates $w_{\rm de}$ very well  especially in the vicinity of present epoch of the Universe, where $a\simeq a_0$. That is why, it has been frequently constrained with observational data in order to study the nature of dynamic dark energy (see \cite{PlanckXXVI} for recent constraints from Planck). However, the CPL parametrization only in terms of scale factor or redshift has been tested with the observational data in bulk of the literature. In recent studies, Akarsu et al. \cite{ozgur12,ozgur14a} investigated a cosmological model based on CPL parametrization of deceleration parameter ($q$) in terms of cosmic time $t$, and observed interesting future evolution of the Universe where it ends in Big Rip. In this model, the authors discussed the Big Rip behavior of the Universe considering the effective cosmic fluid but did not explore the dark energy source responsible for the Big Rip end of the Universe. It is worth noting that  CPL parametrization of deceleration parameter  in $t$ leads to CPL parametrization in $t$ for the EoS parameter ($w$) of the effective cosmic fluid in general relativity in the framework of spatially flat Roberson-Walker (RW) spacetime via the relation $q=-\frac{\ddot{a}a}{\dot{a}^2}=\frac{1+3w}{2}$,
where an over dot denotes derivative with respect to time
$t$. In this study, we shall begin with the CPL parametrization in $t$ for the EoS parameter of the effective cosmic fluid as we are mainly interested to explore the dynamics of the Universe, and in particular the behavior of dynamical dark energy that leads to Big Rip end of the Universe. The work is organized as follows. In Section \ref{sec2}, we give the background equations and the model. In Section \ref{sec3}, we test the success of the model with the latest observational data in contrast with the $\Lambda$CDM model. In Section \ref{sec4}, we explore the nature of dark energy responsible for the Big Rip end of the Universe in the model. We give concluding remarks in Section \ref{sec5}.

\section{Background equations and the model}\label{sec2}
In a Universe filled with a fluid characterized by the EoS parameter $ w_{\rm eff} = p_{\rm eff}/
\rho_{\rm eff}$, the Einstein's field equations (in relativistic units $8\pi G=c=1$) in the framework of spatially flat and homogeneous RW spacetime read as

\begin{equation}\label{eq1}
3H^2=\rho_{\rm eff},
\end{equation}
\begin{equation}\label{eq2}
2\dot{H}+3H^2=- p_{\rm eff},
\end{equation}
where $H=\dot{a}/a$ is the Hubble parameter; $p_{\rm eff}$ and $\rho_{\rm eff}$ are respectively the pressure and energy density of the effective cosmic fluid.

Following the study \cite{ozgur14a}, we consider CPL parametrization for the EoS parameter $w_{\rm eff}$ of the effective fluid in terms of time $t$, that is,
\begin{equation}\label{eq3}
w_{\rm eff}(t)=w_0+w_1\left(1-\frac{t}{t_{0}}\right),
\end{equation}
where $w_0$ is the present value of the effective EoS parameter; $t_{0}$
is the present time or age of the Universe and $w_1$ is a
constant. The above form of $w_{\rm eff}(t)$ is well motivated as discussed in the introduction.

Using \eqref{eq1}-\eqref{eq3} into the relation $w_{\rm eff}(t)=p_{\rm eff}/\rho_{\rm eff} $ and solving, we
obtain the following solution for the scale factor:
\begin{equation}
a(t)=a_{1} \left[\dfrac{w_1 t}{2(1+w_0+w_1)t_0-w_1t}\right]^{\frac{2}{3(1+w_0+w_1)}},
\end{equation}
where $a_{1}$ is a constant of integration, the first constant of integration being assumed zero.

The Hubble parameter $H$, deceleration parameter $q$ and the jerk parameter $j=\frac{\dddot{a}}{aH^3}$ are respectively given by
\begin{equation}
H(t)=\frac{4t_{0}}{3t[2(1+w_0+w_1)t_0-w_1t]},
\end{equation}

\begin{equation}
q(t)=\frac{1}{2}+\frac{3}{2}\left[w_0+w_1\left(1-\frac{t}{t_{0}}\right)\right],
\end{equation}

\begin{equation}
j(t)=\frac{27 w_1^2t^2}{8t_0^2}-\frac{9w_1W t}{4t_0}+\frac{W(1+W)}{2},
\end{equation}
where $W=1+3w_0+3w_1$.

The effective pressure and effective energy density are, respectively
obtained as
\begin{equation}
p_{\rm eff}(t)=\frac{16t_{0}[-w_1t+(w_0+w_1)t_{0}]}{3t^2[2(1+w_0+w_1)t_0-w_1t]^2},
\end{equation}
\begin{equation}
\rho_{\rm eff}(t)=\frac{16t_{0}^2}{3t^2[2(1+w_0+w_1)t_0-w_1t]^2}.
\end{equation}

It is observed that the parameters $a$, $H$, $p$ and $\rho$ diverge to
infinity at two distinct epochs $t=0$ and $t=2(1+w_0+w_1)t_0/w_1$. On
the other hand, the parameters $w$, $q$ and $j$ are well behaved from
former to the later epoch. It implies that in this model, the  Universe
begins with Big Bang at $t=0$ and ends in Big Rip at $t_{\rm
  BR}=2(1+w_0+w_1)t_0/w_1$. The de Sitter time $t_{\rm dS}$, which is
the solution of $q(t_{\rm dS})=-1$, is $t_{\rm
  dS}=(1+w_0+w_1)t_0/w_1$. One may notice that $t_{\rm
  dS}=\frac{1}{2}t_{\rm BR}$. Thus,  the
Universe achieves de Sitter phase at the end of its half life.
The fluid considered here, hence leads to super accelerated expansion 
of the Universe leading to Big-Rip. Thus, Big Rip is an essential feature of the model under consideration. Hereafter, we shall refer it to as BR model.

\section{Observational constraints and comparison with $\Lambda$CDM model}\label{sec3}

In order to confront the BR model with observational data, we utilize $a(t_{0})=1$ and $a(t)=1/(1+z)$, where $z$ stands for redshift. 
The relationship between the cosmic time and redshift in the BR model is given by
\begin{equation}
t=\frac{2(1+w_0+w_1)t_{0}}{w_1+(2+2w_0+w_1)(1+z)^{\frac{3(1+w_0+w_1)}{2}}}.
\label{eq:tzbr}
\end{equation}
The corresponding Hubble parameter quantifying the evolution dynamics reads as

\begin{equation}
H(z)=\frac{H_{0}}{4m^2 } \left[w_1(1+z)^{-\frac{3m}{4}}+ (2m-w_1)(1+z)^{\frac{3m}{4}}\right]^2,
\end{equation}
where $m=1+w_0+w_1$.
We see that the BR model has three free parameters namely $H_0$, $w_0$ and $w_1$. We constrain the parameter space $( H_0,  w_0, w_1)$ of the BR model using 25 observational H(z) points and 580 SN Ia points (see \cite{ozgur14a} and references therein for details) using the Markov Chain Monte Carlo (MCMC) method coded in the publicly available package {\bf cosmoMC} \cite{ref:MCMC}. We use the methodology as described in \cite{ozgur14a}. The results are shown in Table \ref{tab:results}, where the mean values of the parameters are given with errors upto 3$\sigma$ level.
\begin{table}[h]
 \caption{Observational constraints on the parameters of BR model from H(z)+SN Ia data.} 
\begin{center}
\begin{tabular}{ll}
\hline\hline Parameter&Mean values with errors\\ \hline \\
$H_0$	& $ 69.973^{ +0.416 +0.844 +1.278  }_{ -0.427 -0.809 -1.101  }$ \\[8pt]
$w_0$& $ -0.758^{ +0.050 +0.099 +0.160  }_{ -0.051 -0.102 -0.144  }$ \\[8pt]
$w_1$& $ 0.909^{ +0.174 +0.287 +0.386  }_{ -0.151 -0.311 -0.500  }$ \\[8pt]
\hline\hline
\end{tabular}
\label{tab:results}
\end{center}
\end{table}

\begin{figure}[htb!]
\centering
\includegraphics[width=9.5cm]{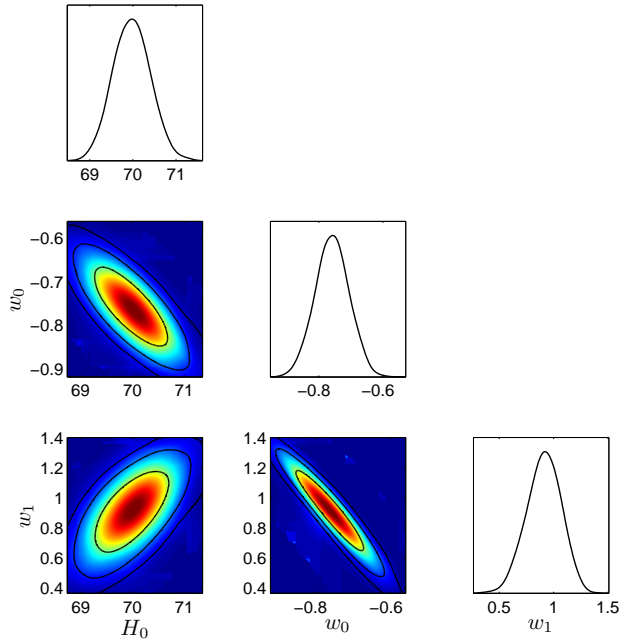}
\caption{1D marginalized distribution on individual parameters and 2D contours with 68.3 \%, 95.4 \% and 99.73 \% confidence limits on parameters of the BR model.}   \label{Fig1}
\end{figure}

\begin{table}[htb!]\centering\small
\caption{Mean values with $1\sigma$ errors of some important cosmological parameters  related to BR and $\Lambda$CDM models.} 
\begin{tabular}{lllll}
\hline\hline Parameters & BR model  & $\Lambda$CDM model\\ \hline\hline\\

$H_0$ (${\rm km}\,s^{-1}\,{\rm Mpc}^{-1}$)  & $69.973^{ +0.416}_{ -0.427}$& $69.777_{-0.318}^{+0.344}$ \\[8pt]

$q_0$ & $-0.637\pm 0.076$ &$-0.572\pm 0.026$ \\[8pt]

$j_{0}$ & $1.600\pm 0.411$ & $1$ \\[8pt]

$\rho_{\rm eff0}\;(10^{-27}$ kg m$^{-3}$) & $9.198\pm 0.109$ & $9.146\pm 0.008$ \\[8pt]
 
$w_{0}$ & $ -0.758_{-0.050}^{+0.051}$ & $-0.714\pm 0.017$\\[8pt]

 $z_{\rm tr}$ & $0.680_{-0.097}^{+0.167}$  & $0.711\pm 0.048$ \\[8pt]
 
 $t_{\rm tr}$ (Gyr) & $7.125\pm 0.652$ & $7.276\pm 0.160$ \\[8pt]
 
 $t_0 $ (Gyr)& $13.379\pm 0.650$ & $13.699 \pm  0.189$  \\[8pt]
 
 $z_{\rm dS}$ & $-0.218\pm{0.096}$  & $-1$ \\[8pt]
 
 $t_{\rm dS}$ (Gyr) & $16.930\pm 0.696$ & $\infty$ \\[8pt]
 
$t_{\rm BR} $ (Gyr)& $33.861\pm 4.138$ & No Big Rip \\[8pt]
 
 \hline
$\chi^2_{min}$ & $575.395$& $575.540$ \\[8pt]
$\chi^2_{min}$/dof & $0.9510$& $0.9513$ \\[8pt]
AIC  & $581.395$  &$579.540$       \\[8pt]
KIC &$584.395 $     & $581.540 $   \\[8pt]
BIC & $594.611$  &$588.351$\\[8pt]
\hline\hline
\end{tabular}
\label{table:BRLCDM}
\end{table}

The 1D marginalized distribution on individual parameters and 2D contours with 68.3 \%, 95.4 \% and 99.73 \% confidence limits are shown in Fig.\ref{Fig1} for the BR model.

Table \ref{table:BRLCDM} shows the values of various parameters pertaining to the BR and $\Lambda$CDM models. We observe that the present values  $H_0$, $q_0$ and $j_0$ of Hubble parameter, deceleration parameter and jerk parameter respectively, which describe kinematics of the current Universe, are in close agreement for the two models. The present values $\rho_{\rm eff0}$ and $w_{\rm eff0}$ of effective energy density and effective EoS parameter respectively, indicate that the two models exhibit similar dynamics of the Universe at the present epoch. The deceleration-acceleration transition redshift $z_{\rm tr}$, transition time $t_{\rm tr}$ and age of the Universe today $t_0$ also agree in the two models . Thus, we see that the two models are almost indistinguishable at the present epoch. However, these exhibit entirely different behavior at late times. For, the $\Lambda$CDM model Universe achieves the de Sitter phase after infinite time and continues to be in the same phase thereafter to expand forever with exponential expansion. On the other hand, the BR model Universe exhibits the de Sitter phase in finite time at $t_{\rm dS}=16.930\pm 0.696$ Gyr ($z_{\rm dS}=-0.218\pm{0.096}$). Thereafter, it expands with super acceleration ($q<-1$) and ends in Big Rip at $t_{\rm BR}=33.861\pm 4.138$ Gyr.  Thus, the BR model offers an entirely different future of the Universe in contrast with the  $\Lambda$CDM model.
\begin{figure}[h]
\centering
{\psfrag{w}[b][b]{$w(z)$}
\psfrag{z}[b][b]{$z$}
\psfrag{d}[b][b]{\textcolor{blue}{$w=-\frac{1}{3}$}}
\psfrag{wa}[b][b]{\textcolor{blue}{$w=-\frac{1}{3}$}}
\includegraphics[width=8.5 cm]{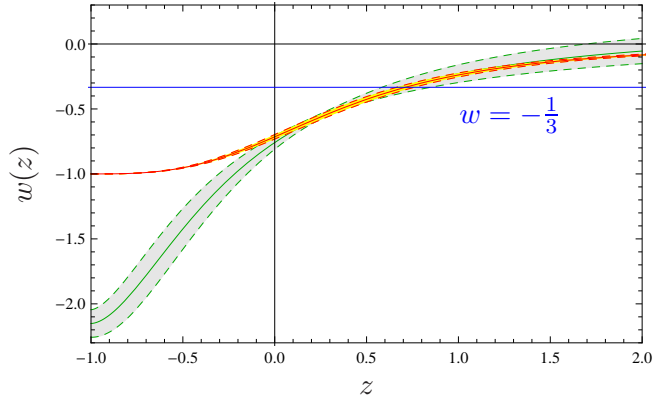} }
\caption{\footnotesize{The effective EoS parameters of BR model (Green curves) and $\Lambda$CDM model (Red curves) are shown versus redshift. The solid Green and Red curves correspond to the mean values of the EoS parameters while the shaded regions between the dotted curves are $1\sigma$ error regions. The Blue colored horizontal line stands for $w=-\frac{1}{3}$.}}
\label{figwz}
\end{figure}

Akaike Information Criterion (AIC), Kullback Information Criterion (KIC) and Bayes Information Criterion (BIC) are the statistical tools, which  are commonly used in modern cosmology for model selection among competing models. These are defined as   (see \cite{Melia2013,ozgur14b} and references therein for details)
\begin{equation}
\text{AIC}=\chi^2+2k,\; \text{KIC}=\chi^2+3k,\;\text{BIC}=\chi^2+k\ln n,
\end{equation}
where $k$ stands for the number of parameters appearing in the model while $n$ is number of data being used in fitting. The values of AIC, BIC and KIC for the BR and $\Lambda$CDM models are displayed in
Table \ref{table:BRLCDM}. The  $\Lambda$CDM model appears with lower values of AIC, KIC and BIC in comparison to the BR model, and thereby is closer to the real model. For testing the relative evidence of better model among the two models under consideration, we find the difference of the values of AIC, KIC and BIC, which read as $\Delta \text{AIC}=1.855$, $\Delta \text{KIC}=2.855$ and  $\Delta \text{BIC}=6.260$. Now considering the thumb rules of AIC, BIC and KIC (see \cite{ozgur14b}), as per AIC the BR and $\Lambda$CDM  models are equally preferable while the values of $\Delta \text{KIC}$ and $\Delta \text{BIC}$ respectively show mild and strong evidences against the BR model. In the cases of KIC and BIC, in fact, the BR model is penalized because it carries one extra free parameter in comparison to the $\Lambda$CDM  model.

In Fig. \ref{figwz}, we show the evolution of the effective EoS parameters of BR model (Green curves) and $\Lambda$CDM model (Red curves) versus redshift, where the shaded regions between the dotted curves are $1\sigma$ error regions while the solid curves correspond to the mean values of the respective parameters. We observe that the effective cosmic fluids in the two models evolve in a similar manner till the present epoch but their evolution curves differ significantly in future, and thereby giving rise to  distinct futures of the Universe, viz., Big Rip in the BR model and de Sitter in the $\Lambda$CDM model.

\section{Matter and dark energy sources in BR model}\label{sec4}
In the previous two sections, we discussed the dynamics of the Universe considering the effective cosmic fluid. Here, we shall isolate the components of the effective fluid, viz., matter (baryonic+CDM) and dark energy. We have neglected the radiation component considering its negligible contribution from the matter dominated epoch onwards. So we have $p_{\rm eff}=p_{\rm m}+p_{\rm de}$ and $\rho_{\rm eff}=\rho_{\rm m}+\rho_{\rm de}$, where $p_{\rm m}$ and $p_{\rm de}$ are pressures while  $\rho_{\rm m}$ and $\rho_{\rm de}$ are energy densities of the matter and dark energy components respectively. Now, equation \eqref{eq1} can be cast into the form
\begin{equation}\label{de_eos}
\Omega_{\rm de}=1-\Omega_{\rm m},
\end{equation}
where $\Omega_{\rm m}=\rho_{\rm m}/(3H^2)$ and  $\Omega_{\rm de}=\rho_{\rm de}/(3H^2)$ are respectively the matter and dark energy density parameters.
The EoS parameter of the matter component reads as $w_{\rm m}=p_{\rm m}/\rho_{\rm m}=0$, matter being pressureless. Further, the energy density of matter varies as $\rho_{\rm m}=\rho_{\rm m0}(1+z)^3$, $\rho_{\rm m0}$ being the value of energy density of matter today. 

With the above considerations into equations \eqref{eq1} and \eqref{eq2}, the EoS parameter $w_{\rm de}=p_{\rm de}/\rho_{\rm de}$ of dark energy component in the BR model can be obtained as
\begin{equation}\label{de_eos}
w_{\rm de}(z)=\frac{2(1+z)HH'-3H^2}{3H^2-3H_0^2\Omega_{\rm m0}(1+z)^3},
\end{equation}
where prime denotes the derivative with respect to $z$, and $\Omega_{\rm m0}=\rho_{\rm m0}/(3H_0^2)$ is the density parameter of matter for the present day Universe. However, the BR model under consideration being a model based on EoS parameter of the effective cosmic fluid, does not provide the explicit contribution of matter. Nevertheless, we need to know the contribution of matter in the BR model in order to study the dynamical behavior of dark energy by using equation \eqref{de_eos}. In this regard, we look for the cosmic microwave background (CMB) shift parameter, which for a spatially flat Universe is defined as (see \cite{Elgaroy07} and references therein)
\begin{equation}\label{SP}
R=\sqrt{\Omega_{\rm m0}}\int_{0}^{z_{dec}}\frac{H_{0}}{H(z)} {\rm d}z,
\end{equation}
where $z_{\rm dec}$ is the redshift of decoupling, usually fixed to 1090.  This parameter is related to the position of the first acoustic peak in the power spectrum of the CMB temperature anisotropies, and is used for simple tests of dark energy models \cite{Elgaroy07}. In the context of the BR model under consideration, the analysis of the shift parameter will not only give us the idea about matter density in the BR model but also let us know about the consistency of BR model at higher redshifts given that it does very well at low redshifts for describing the evolution of the Universe. 

Using the values of the parameters given in Table \ref{tab:results}, we find $R/\sqrt{\Omega_{\rm m0}}=2.80\pm 0.47$ for the BR model while for the $\Lambda$CDM model, we find $R/\sqrt{\Omega_{\rm m0}}=3.25\pm 0.07$. One may see the overlap of the values of the ratio $R/\sqrt{\Omega_{\rm m0}}$ within 1$\sigma$ error region for the two models. For further analysis, we have the option of fixing either the value of $\Omega_{\rm m0}$ or $R$. First, we shall fix the value of $R$ from some reliable source. The  $\Lambda$CDM model, when constrained with the observational data from $H(z)$ and SN Ia compilations used in this study, gives $\Omega_{\rm m0}=0.285_{-0.017}^{+0.018}$. Further, we find the shift parameter for the $\Lambda$CDM model as $R_{\Lambda{\rm CDM}}=1.741\pm0.042$, which is consistent with the measured value $R_{\rm Planck} =1.744\pm 0.011$  in the Planck experiment \cite{PlanckXXVI,King13}. Now, choosing   the values of $R$ from Planck experiment and the mean values of other parameters given in Table \ref{tab:results}, we find $\Omega_{\rm m0}=0.387\pm0.130$ for the BR model. This shows that the BR model requires higher mean value of the matter density parameter in contrast with the $\Lambda$CDM model. However, within 1$\sigma$ error region it accommodates the range of values of $\Omega_{\rm m0}$ given by $\Lambda$CDM model. On the other hand, if we choose the values of  $\Omega_{\rm m0}$ as obtained in the $\Lambda$CDM model, we find $R_{\rm } =1.495\pm0.255$. Thus, the BR model shows the consistency with the CMB shift parameter within $1\sigma$ error region, and thereby does reasonably well at higher redshifts too. 

\begin{figure}[htb!]
\centering
\subfigure[]
{\psfrag{w}[b][b]{$w_{\rm de}(z)$}
\psfrag{z}[b][b]{$z$}
\includegraphics[width=8cm]{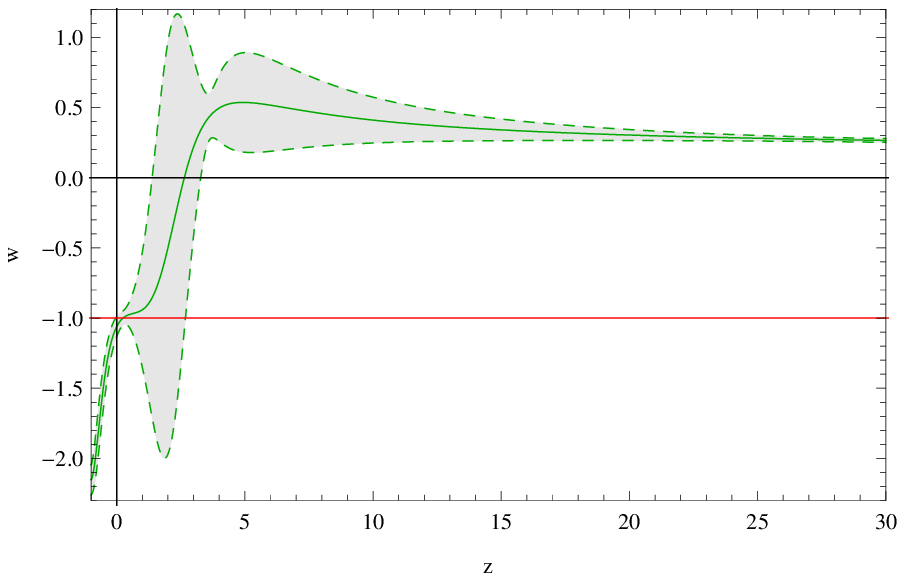} \label{figwdez30}}
\hspace{0.1cm}
\subfigure[]{
\psfrag{w}[b][b]{$w_{\rm de}(z)$}
\psfrag{z}[b][b]{$z$}
\includegraphics[width=8 cm]{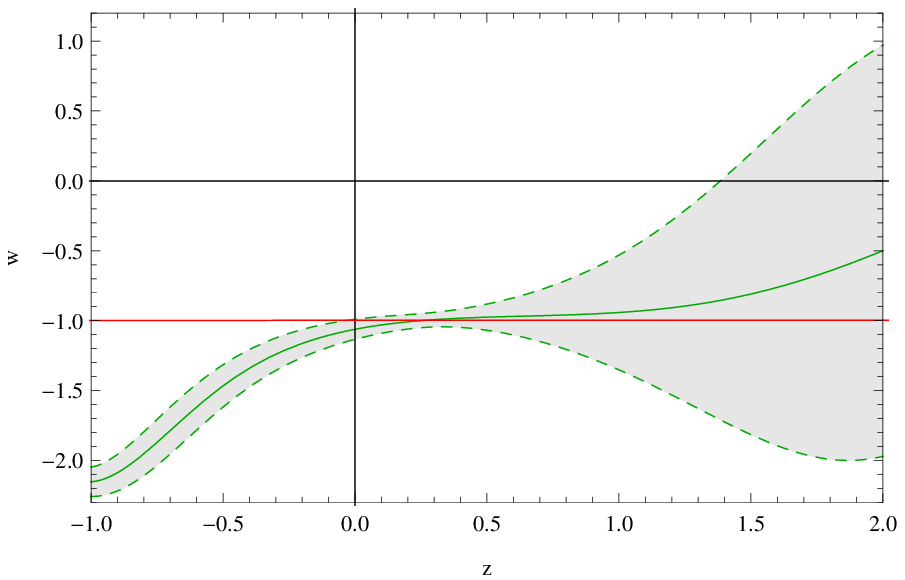} \label{figwdez}}
\caption{\footnotesize{(a) The dark energy EoS parameters of BR model (Green curves) and $\Lambda$CDM model (horizontal Red line) are shown versus redshift. The solid Green curve corresponds to the mean value of the dark energy EoS parameter in the BR 	model while the shaded region between the dotted Green curves is $1\sigma$ error region. (b) A closer view of the left panel figure in the redshift range $[-1,2]$ is displayed.}}
\end{figure}

Since the BR model is in very close agreement with the $\Lambda$CDM model at the present epoch (see Table \ref{table:BRLCDM}), it is reasonable to use the value of $\Omega_{\rm m0}=0.285_{-0.017}^{+0.018}$ as suggested by the $\Lambda$CDM model. So for extracting the behavior of the dynamical dark energy in the BR model, we plug in the mean value $\Omega_{\rm m0}=0.285$ into equation \eqref{de_eos} and, thereby we find $w_{\rm de0}=-1.061\pm0.071$ using the constraint results given in Table \ref{tab:results}. So the present value of the EoS parameter of dark energy in the BR model coincides the EoS parameter $w_{\Lambda}=-1$ of the vacuum energy in the  $\Lambda$CDM model. We now analyze the dynamical behavior of dark energy source in the BR model by observing the variation of its EoS parameter $w_{\rm de}$ vs $z$ as shown in Fig. \ref{figwdez30}. We see that there is a steep downfall of the values of $w_{\rm de}$ after the redshift around $z=5$. This dramatic and sudden change in the behavior of dark energy source in the BR model is indicated by the larger error region around the mean value curve of $w_{\rm de}$ around the redshift $z=5$. It is the phase when the dark energy source gears up to dominate the matter source, and finally shifts the expansion of the Universe from deceleration to acceleration around the redshift $z_{\rm tr}=0.68$ (see Table \ref{table:BRLCDM}). Thereafter, the Universe expands with acceleration. It is interesting to observe that from the onset of acceleration ($z=0.68$) till the present epoch ($z=0$), the $w_{\rm de}$ curve tends to be flat with the  $w_{\Lambda}=-1$ line, that is, $w_{\rm de}\simeq-1$, as may be seen from Fig. \ref{figwdez}, which gives a closer view of the Fig. \ref{figwdez30} in the redshift range $[-1,2]$. So the dark energy in BR model behaves like cosmological constant from the onset of acceleration till the present epoch. Also, it seems to stabilize as cosmological constant during this regime. However, this is not the case in future. In the redshift range $[-1,0]$, one may see that $w_{\rm de}$ curve falls down to values less than $-1$ crossing the phantom divide line $w_{\rm de}=-1$. This shows that in future, the dark energy source of BR model evolves in the phantom region, and pushes the expansion of the Universe to super acceleration. The super accelerating Universe, finally, ends with Big Rip at a time around $33.86$ Gyr (see Table \ref{table:BRLCDM}).

\section{Concluding remarks}\label{sec5}

The BR model describes the Universe from the matter-dominated phase to the recent dark energy dominated phase in line with the $\Lambda$CDM model but offers a Big Rip end of the Universe contrary to the everlasting de Sitter expansion of the Universe in the $\Lambda$CDM model. It fits the observational data from the $H(z)$ and SN Ia compilations matching the success of the $\Lambda$CDM model. The analysis further reveals that both the models describe almost identical evolution of the Universe from the onset of acceleration till the present epoch, in particular.  In future, the $\Lambda$CDM model evolves to the de Sitter phase with constant vacuum energy density. On the other hand, the dark energy in the BR model is dynamical in nature. Interestingly, it mimics the cosmological constant behavior in the vicinity of present epoch, and exhibits phantom behavior in future. Due to the dominance of this dark energy, the Universe has finite life time in the BR model and it ends in Big Rip. 

Usually, solutions of the Einstein's field equations in general relativity are found and analyzed
for different epochs, that is, for inflationary
phase ($p_{\rm eff}\simeq -\rho_{\rm eff}$), radiation-dominated phase ($p_{\rm eff}\simeq\rho_{\rm eff}/3$) and matter-dominated phase ($p_{\rm eff}\simeq 0$). However, considering equation of state (EoS) parameter ($w_{\rm eff}=p_{\rm eff}/\rho_{\rm eff}$) for the effective fluid, unified solutions for these epochs are also presented by some authors in the literature. For
instance, Israelit and Rosen \cite{14,15} considered a phenomenological form of the effective EoS parameter, and described the evolution of the Universe from pre-matter period
 to the radiation-dominated phase,  and then
radiation to matter-dominated period. 
Similarly, a phenomenological form of
effective EoS parameter was suggested by Carvalho \cite{16} to describe a unified evolution of the
Universe from the inflationary phase to the radiation-dominated phase.
However, the BR model investigated in this paper is based on the naturally motivated CPL parametrization for EoS parameter of the effective cosmic fluid.  We would like to emphasize that the CPL parametrization is originated from the Taylor expansion of the EoS parameter. So it does not strictly belong to the class of phenomenological parameterizations of the EoS parameter. In fact, it is a precise measure of the real EoS parameter upto the first order terms in $t$.

As a final note for the prospective readers we would like to mention that the BR model does very well for describing the evolution of the Universe at low redshifts, and describes the Big Rip future of the Universe as intended in this study. However, it does well at higher redshifts too as we observed from the CMB shift parameter analysis. Probably, one more term is required in the Taylor approximation of $w_{\rm eff}(t)$ for much better performance of BR model at higher redshifts. But it will bring in an additional parameter to the model, and  may introduce difficulty in analytical solution of the model. Also, the model generated by considering one extra term may not describe a Universe exhibiting the Big Rip feature essentially. Nevertheless, this idea deserves attention for further investigation.  Next, it may be worthwhile to investigate a possible generalization of the BR model by considering a coupling between dark matter and dark energy (see \cite{Amen04}). It would be interesting to know whether (and how) the results of this paper might change in the presence of a direct coupling between the dark matter and dark energy components of the cosmic fluid.

\subsection*{Acknowledgments}
The author is thankful to \"{O}zg\"{u}r Akarsu for fruitful discussions. The author is supported by the Department of Science and Technology, India under project No. SR/FTP/PS-102/2011.

\end{document}